\documentclass{article}

\usepackage{PRIMEarxiv}

\usepackage[utf8]{inputenc} % allow utf-8 input
\usepackage[T1]{fontenc}    % use 8-bit T1 fonts
\usepackage{hyperref}       % hyperlinks
\usepackage{url}            % simple URL typesetting
\usepackage{booktabs}       % professional-quality tables
\usepackage{amsfonts}       % blackboard math symbols
\usepackage{nicefrac}       % compact symbols for 1/2, etc.
\usepackage{microtype}      % microtypography
\usepackage{lipsum}
\usepackage{fancyhdr}       % header
\usepackage{graphicx}       % graphics
\usepackage{amsmath}
\usepackage{amssymb}
\usepackage{booktabs}
\usepackage{subcaption}

\usepackage{algorithm}
\usepackage{algorithmic}
\usepackage[T1]{fontenc}
\graphicspath{{media/}}     % organize your images and other figures under media/ folder

\begin{document}

\title{Single-Image Based Deep Learning for Precise Atomic Defects Identification}

\author{Kangshu Li$^{1, \#}$, Xiaocang Han$^{1, \#}$, Yanhui Hong$^{2, *}$, Yuan Meng$^{1}$, Xiang Chen$^{1}$, Junxian Li$^{1}$, Jing-Yang You$^{3}$, \\
\textbf{Lin Yao$^{2}$, Wenchao Hu$^{1}$, Zhiyi Xia$^{2}$, Guolin Ke$^{2}$, Linfeng Zhang$^{2,4}$, Jin Zhang$^{1}$, Xiaoxu Zhao$^{1, 4, *}$} \\
\\
$^1$ School of Materials Science and Engineering, Peking University, Beijing 100871, China \\
$^2$ DP Technology, Beijing, 100080, China \\
$^3$ Department of Physics, National University of Singapore, 117551, Singapore \\
$^4$ AI for Science Institute, Beijing 100084, China \\
}

\maketitle

\begin{abstract}
Defect engineering has been profoundly employed to confer desirable functionality to materials that pristine lattices inherently lack. Although single atomic-resolution scanning transmission electron microscopy (STEM) images are widely accessible for defect engineering, harnessing atomic-scale images containing various defects through traditional image analysis methods is hindered by random noise and human bias. Yet the rise of deep learning (DL) offering an alternative approach, its widespread application is primarily restricted by the need for large amounts of training data with labeled ground truth. In this study, we propose a two-stage method to address the problems of high annotation cost and image noise in the detection of atomic defects in monolayer 2D materials. In the first stage, to tackle the issue of data scarcity, we employ a two-state transformation network based on U-GAT-IT for adding realistic noise to simulated images with pre-located ground truth labels, thereby infinitely expanding the training dataset. In the second stage, atomic defects in monolayer 2D materials are effectively detected with high accuracy using U-Net models trained with the data generated in the first stage, avoiding random noise and human bias issues. In both stages, we utilize segmented unit-cell-level images (usually one cell per image) to simplify the model's task and enhance its accuracy. Our results demonstrate that not only sulfur vacancies, we are also able to visualize oxygen dopants in monolayer $MoS_2$, which are usually overwhelmed by random background noise. As the training was based on a few segmented unit-cell-level realistic images, this method can be readily extended to other 2D materials. Therefore, our results outline novel ways to train the model with minimized datasets, offering great opportunities to fully exploit the power of machine learning (ML) applicable to a broad materials science community. 
\end{abstract}

\section{Introduction}

Defects, intricately woven into the very fabric of crystalline materials, serve as the unseen architects, bestowing upon them a kaleidoscope of captivating electronic, optical, chemical, magnetic, and mechanical fuctions~\cite{1,2}. Sulfur vacancies, for example, a prevalent point defect in $MoS_2$, are harnessed to introduce fascinating phenomena—from the emergence of in-gap states~\cite{3}, and the transition from p-type to n-type semiconductor~\cite{4,5}, to the enhancement of photoluminescence~\cite{6}, and the supreme hydrogen evolution reaction activity~\cite{7}. Thanks to the advancement in aberration-corrected STEM, revealing the intricate tapestry of defect structures at the atom scale is routinely accessible in real space, encompassing vacancies, substitution, antisites, dislocation cores, edges, and grain boundaries~\cite{8,9,10,11}. Utilizing Z-contrast high-angle annual dark-field STEM (HAADF-STEM) imaging, where the intensity is proportional to atomic number ($Z^{1.6-1.9}$ )\cite{8}, allows for the categorization and mapping of these defects~\cite{zhou2013intrinsic}. However, traditional histogram-based methods of image intensity analysis are highly susceptible to human errors, due to the presence of detector noise, aberrations, scanning distortions, sample drift, surface contamination, alignment errors, and filtering artifacts, all of which cause variations in contrast~\cite{jones2015smart,krause2016effects,wang2018correcting}. Such methods typically require laborious manual processing and quantification, leading to both low throughout and low-confidence predications~\cite{8,12}. Hence, establishing an automated method compatible with atomic precision and high throughput for defect analysis is of significant demand.

Machine learning (ML), particularly deep learning (DL), revolutionizing the field of image recognition, has been widely applied in medical imaging diagnosis~\cite{13}, autonomous vehicles~\cite{14}, facial recognition systems~\cite{15,16}, and various domains of science~\cite{17,18}. The exploitation of ML in STEM image analysis has emerged as a new platform for delving into the atomic defect universe with high-throughput structure analysis and veracity~\cite{19,20,21,22,23} and uncovering and projecting the deep correlations of STEM images. For example, by utilizing the DL network, vacancy sites~\cite{24}, dopants, a reversible symmetry switching analysis of a single atom, and point defect dynamics can be accurately identified and quantified~\cite{25,26}. Excitingly, the ability to reveal radial strain field oscillations around vacancies has become a reality, no longer just a prophecy. Owing to the assistance of DL-enabled class averaging, an extremely high precision of 0.2 picometers has been realized~\cite{27}. 

Despite the great promise of ML, there are still many challenges to overcome\cite{zhou2017machine,kumar2017data}. First, the range of image quality that the network can reliably handle is limited. Multi-variable experimental environment is prevalent each time the image is collected, only carefully chosen images are processed with the network. Second, a large amount of high-quality training data paired with the ground truth is required for supervised learning methods. Notably, the simulated data inherently including ground truth~\cite{27,28,29,30,31,32,33,34} is typically employed to train models. However, simulated images deviate non-trivially from experimental images, rendering the constructed models inaccurately applicable to real STEM images. To bridge the gap between simulated and experimental images, manually added noise~\cite{27,29,35}and noise enhancement deep learning models~\cite{36} are employed for image transformation, which needs a large dataset to yield enough processed simulations for further defect identification models~\cite{24}. Hence, it is not surprising that defect identification models perform suboptimal in one specific experimental image, due to the inability of a large training set to accurately depict each image's unique noise~\cite{20,21,22,23}, thus limiting generalizability and further accuracy improvement in defect identification. These factors make it challenging to build an ML model that is highly automated, self-adaptive, and covers most of the experimental variations with high-accuracy and precision, especially in systems where defects possess ambiguous intensity differences.

Here, we use only one single experimental image as the training set for our noise augment model, enabling the creation of realistic simulated images that better represent that specific experimental image. Consequently, this enhances the accuracy of the defect identification model trained with this dataset in defect identification in that experimental image. We create this dynamically evolved DL model trained by segmenting both experimental and simulated images into primitive units, to demonstrate its utility for the identification of chalcogen vacancies, especially for sulfur vacancy, and expanding the range of identifiable defect types to oxygen atom doping in monolayer $MoS_2$. An unsupervised generative attentional network with adaptive layer instance normalization for image-to-image translation (U-GAT-IT)~\cite{37}, a kind of Generative Adversarial Networks (\textbf{GAN}), was constructed to generate simulated motifs similar to experimental images, which provides a high-quality training set upon stitching for defect identification tasks using a U-Net based on Fully Convolutional Network (\textbf{FCN})~\cite{25,31}. 

\section{Results and Discussion}
Due to the absence of precise visual discrimination standards, identifying defects in STEM images via human endeavors is challenging to achieve a common ground truth. To reduce the variations in chalcogen defect identifications, seven researchers with a few years of STEM experience are invited to label defects on twelve different STEM images, including three high-quality and three low-quality images for $MoS_2$ and $WSe_2$, respectively. It can be seen that huge disparities were concluded regarding the chalcogen defect identification. In particular, the variations are considerably larger in $MoS_2$, presumably due to the weak contrast of S atoms. The causes of that conclusion are attributed to random scanning distortions and carbon contamination in the image, which lead to differences in contrast and shapes between different atom blobs, particularly when imaging conditions vary, resulting in significant differences in image quality. Consequently, there is no consistent and universal standard suiting for each single image, leading to substantial subjective influence on each researchers annotations. Hence, developing a single image-based DL approach is necessary to achieve high accuracy in defect identification.

\begin{figure}[htb]
		\centering
            \includegraphics[width=0.98\textwidth]{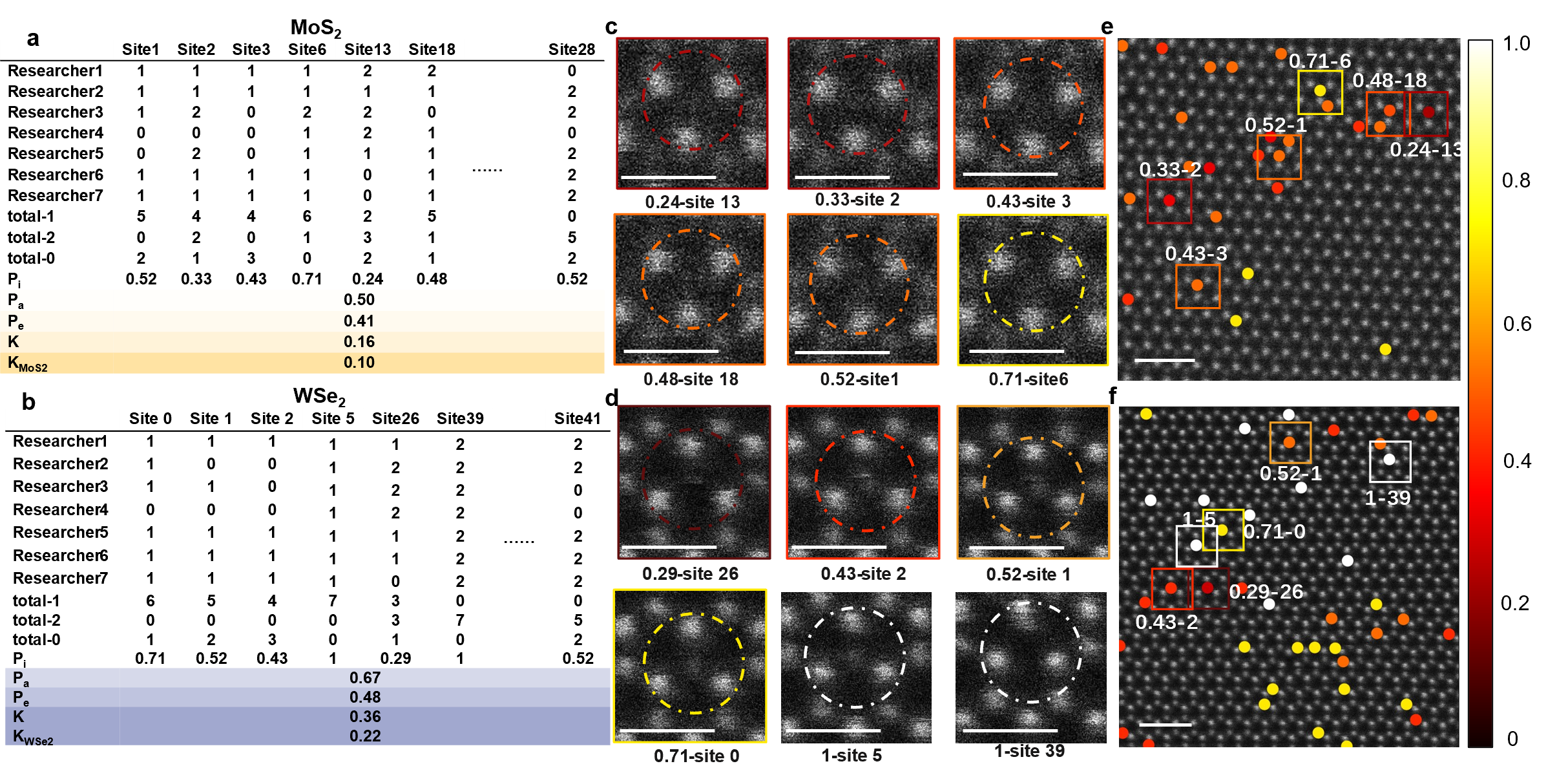}
		\caption{\textbf{Quantitative measurements of the inconsistency of defect identification labeled manually in $MoS_2$ and $WSe_2$.} a, b) Fleiss' Kappa statistic tables of $MoS_2$ and $WSe_2$: the evaluations by different researchers for each potential defect site and the consistency statistics across multiple areas, where ‘0’, ‘1’, ‘2’ represents chalcogen dimmer, a single chalcogen vacancy and a dichalcogen vacancy, respectively. Magnified images of six different sites with distinct Pi values are displayed in the $MoS_2$-Table (c), and $WSe_2$-Table (d). Defect locations are highlighted by circles, with corresponding K values noted below the image. (e, f) Heat maps of the K values for potential defect sites. The consistency of annotations is indicated by the varying colors of the marked circles, representing K, with a color-to-K correlation scale displayed on the right. Scale bars: 0.5 nm in (c, d); 1 nm in (e, f). }
		\label{fig:1}
\end{figure}

To quantify the defects decision variations, the Fleiss' Kappa statistic (K), a widely recognized metric for gauging rater agreement~\cite{38,39}, served as the quantitative evaluation metric. A higher K value, indicating a larger proportion of researchers making the same choice, suggests greater reliability in manual defect analysis. Figure~\ref{fig:1}a and 1b illustrate the average K for all $MoS_2$ images ($k_{MoS_2} = 0.1$) is markedly lower compared to $WSe_2$ ($k_{WSe_2} = 0.22$). Therefore, there is no common ground truth for defect identification in 2D materials via conventional human endeavors, particularly when the image quality is poor or the chalcogen atom is weak.

In pursuit of the highest precision of defect identification, we aim to dynamically evolve DL models using a single experimental image, which is akin to tailoring a personalized defect identification system suitable for each distinct experimental scenario. One of the foremost challenges in achieving the dynamic evolving of ML models lies in reconciling the contradiction between the substantial volume of experimental data required for conventional training DL models and the need for fine-tuned model parameters specific to the individual experimental image. 

To tackle the absence of sufficient training data provided by only a single experimental image, we segment both the single experimental image and corresponding simulated data into the smallest primitive units to serve as the training set for the noise augment model (U-GAT-IT). The U-GAT-IT-processed unit simulated images, used as the training set for the defect identification model (U-Net), not only accurately describe this experimental image but also increase the diversity of the training set through random assembly, meeting the demand for a large quantity of training data. This approach harmonizes the dual requirements of extensive data for model training and the specificity needed for individual image analysis, thus enhancing the applicability and precision of the model for defect identification. 

\begin{figure}[htb]
		\centering
            \includegraphics[width=0.98\textwidth]{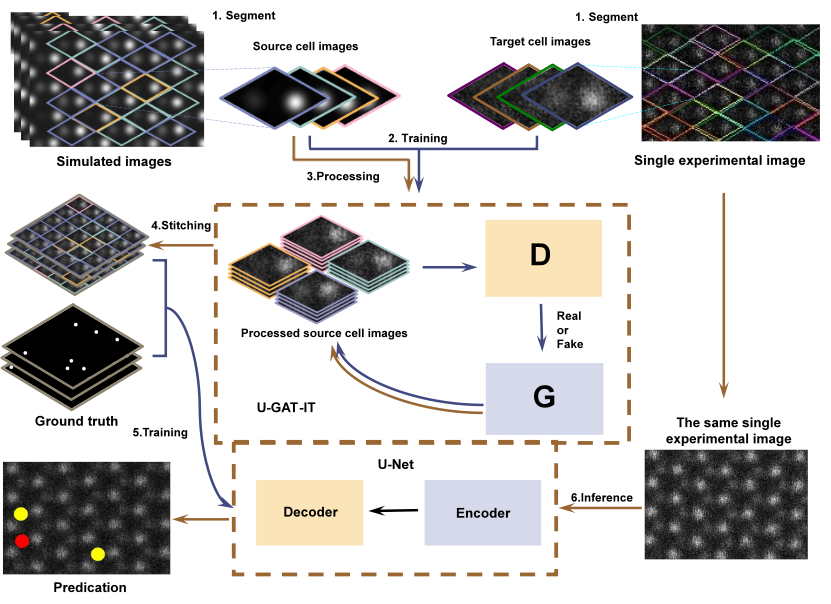}
		\caption{\textbf{The workflow of detecting single-atom defects based on a single image.} The single experimental STEM image to be identified and its corresponding simulated image are segmented before being processed through the identification system, \textit{i.e.}, U-GAT-IT and U-Net models. The purple arrows represent the fine-tuned training process of the two models, whereas the brown arrows indicate the operational workflow. G and D indicate generator and discriminator, respectively.}
		\label{fig:2}
\end{figure}

This precise defect identification method based on a single experimental image is characterized by the segmentation and assembly steps demonstrated in the first and third steps (Figure~\ref{fig:2}). The main six steps are also presented in sequence (Details are described in Methods): (1) segmentation of an experimental image and corresponding simulations, (2) training U-GAT-IT with experimental and simulated unit images, (3) applying the U-GAT-IT to process unit simulated images for noise augmentation, (4) reconstructing processed simulated images, (5) U-Net training with stitched simulated images, (6) employing U-Net to identify defects in experimental image. This method involves cropping a single experimental image into primitive units containing only two atomic columns, as this is the only way to ensure that the same unit simulated images can extract noises from as many different areas of the experimental image as possible. This is the first step in expanding from a single simulated image to n images, n is determined by the size of the selected STEM image. Furthermore, assembling sixteen processed realistic unit simulated images through image stitching randomly serves to expand a single U-Net training image into N images. These two steps effectively address the challenge induced by insufficient amount of variability in training data for U-GAT-IT and U-Net models while maintaining the specific noise source, which is the biggest challenge introduced when relying on only a single image as a basis. 

\begin{figure}[htb]
		\centering
            \includegraphics[width=0.98\textwidth]{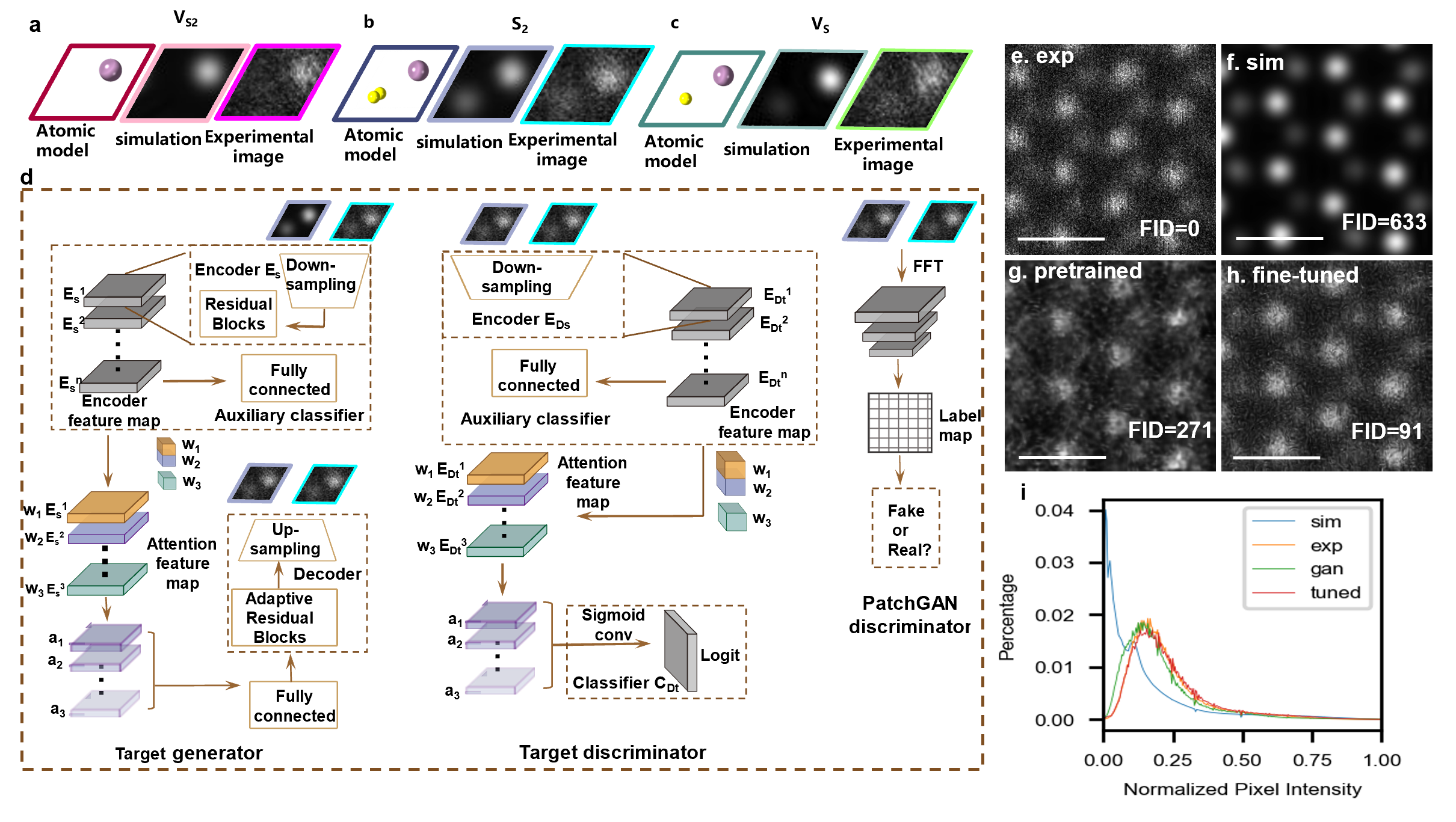}
		\caption{\textbf{Input, architecture, and evaluation metrics of fine-tuned U-GAT-IT.} a-c) Atomic model, simulated and experimental STEM images of three types of unit cells in monolayer $MoS_2$: $V_S$, $S_2$, $V_{S2}$, representing single sulfur vacancy, no vacancy and disulfide vacancy respectively.d) Schematic architecture of a U-GAT-IT containing one generator and two discriminators. e-h) Experimental, simulated, pretrained-U-GAT-IT-processed, and fine-tuned-U-GAT-IT processed images. The FID scores are marked the FID scores are marked in the image. i) The Kullback–Leibler (KL) divergence of $MoS_2$ images generated from (e-h). Scale bars: 0.5 nm in (e-h). }
		\label{fig:3}
\end{figure}

To generate realistic unit simulated images, we apply an unsupervised generative adversarial network (GAN), U-GAT-IT, which integrates a new attention module~\cite{37} and an FFT discriminator, as shown in Figure~\ref{fig:3}d. Through utilizing these attention maps in all the generators and discriminators, our model is guided to focus more on the distinct areas between source and target, in the context of this work, they are referred to the imperfections existing in the experimental images, thereby enhancing the transformation of shapes. In this noise augmentation algorithm comprising one generator and two discriminators, we have advanced its architecture by incorporating an FFT discriminator tailored to its application in transforming between simulated images and STEM images. This approach ensures that the generated images are not only visually similar to the real images but also share similar frequency-domain characteristics~\cite{24,40,41}, which is crucial for accurate and realistic simulations of STEM. 

The performance of the noise augment model is significant because it largely determines whether this method can further enhance the precision of defect identification. We choose experimental images of monolayer $MoS_2$ as testing subjects, where vacancies are challenging to identify, thereby enhancing the reference value of the following evaluation data. All possible intrinsic unit images of it are illustrated in Figure~\ref{fig:3}a. In Figure~\ref{fig:3}e-h, both the pre-trained U-GAT-IT processed image (Figure ~\ref{fig:3}g) and fine-tuned U-GAT-IT processed image (Figure~\ref{fig:3}i) exhibit higher visual similarity to real experimental image (Figure~\ref{fig:3}e) than the simulation (Figur~\ref{fig:3}f). To quantitively evacuate their similarities, we calculate two metrics extensively used to evaluate the image quality, the Fréchet Inception Distance (FID)~\cite{42} and Kullback–Leibler (KL)~\cite{43} divergences between the experimental image and the other three images shown in Figure~\ref{fig:3}. A lower score indicates a greater similarity between the two image sets, or in other words, their statistical properties are more alike. The Kullback-Leibler divergence (KL divergence), also known as relative entropy, is a mathematical metric for quantifying the difference between two probability distributions. The simulated image (FID = 396.32) with the lowest similarity, unsurprisingly, possesses the highest FID value. The fine-tuned U-GAT-IT processed image (FID=91) shows the lowest value, which is also supported by the calculation of the , demonstrating the initial success of our method in training the noise enhancement model.

\begin{figure}[htb]
		\centering
            \includegraphics[width=0.98\textwidth]{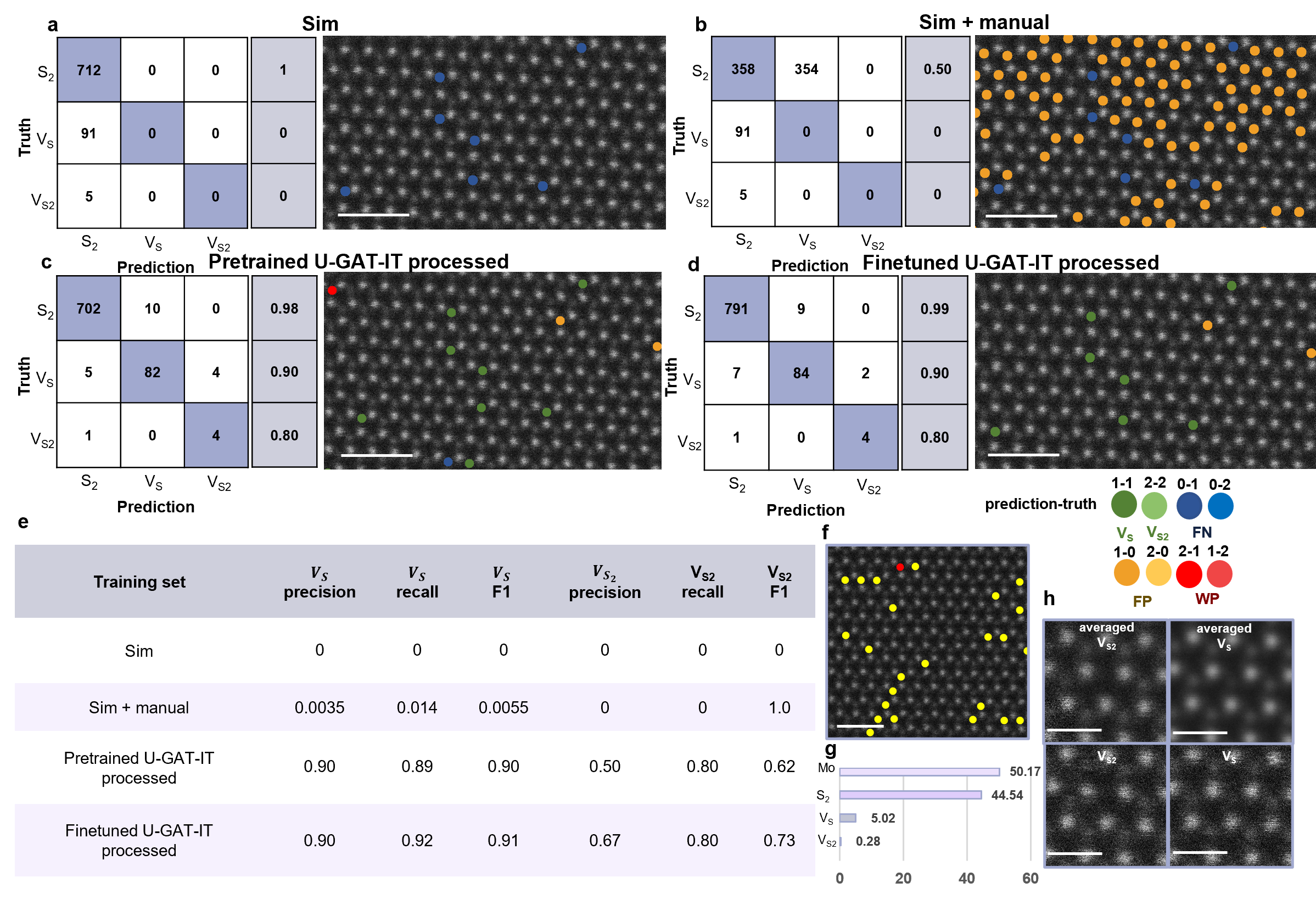}
		\caption{\textbf{Evaluation of U-Net trained by different training sets.} a-d) Confusion matrix obtained by matching the ground truth on the “${MoS_2}-1$” experimental image and the predictions identified by U-Net trained on simulated images (a), simulated images with manual noises (b), pre-trained U-GAT-IT processed simulated images (c) and fine-tuned U-GAT-IT processed simulated image (d). On the right is the prediction result compared to the ground truth, with the color scale located below the image. FN, WP, and FP are the abbreviations of false negative, wrong positive, and false positive, respectively. (e) The table shows six evacuation metrics of U-Net performance with different training sets: F1 scores. recall, precision based on $V_S$ and $V_{S2}$ respectively. (f) Experimental $MoS_2$ image with highlighted defects prediction, red spots, and bright yellow spots represent $V_{S2}$, $V_S$ respectively. (g) Site quantification for composition as an optional function on the website. (h) Class-averaged cropped images of and $V_S$. Scale bars: 1 nm in (a-d, f); 0.5 nm in (h).}
		\label{fig:4}
\end{figure}

Utilizing the same experimental image for testing, we conduct a lateral comparison of defect identification methods by switching between different training sets.The different performance of U-Net models trained with distinct datasets highlights the superiority of this method, as illustrated in Figure~\ref{fig:4}a-d.. This noise augment model endeavors to generate simulated images that closely resemble the experimental images, aiming to enhance the quality of the training set for the defect identification model. The U-Net models are applied based on the FCN algorithm as a defect identification model. 

Subsequent enhancement of defect identification precision, we can qualitatively and quantitatively observe, through the confusion matrix and accuracy plots, that the training set, augmented with noise using U-GAT-IT, becomes more similar to the experimental images. As shown in Figure~\ref{fig:4}e, precision, recall, and F1 scores are chosen to evaluate the trained U-Nets, the U-Net trained with fine-tuned U-GAT-IT-processed images ($V_{s2}  F1 = 0.73$) perform obviously better than the one trained with pre-trained U-GAT-IT-processed simulated images ($V_{s2} F1 = 0.62$). Notably, in most research, the evaluation of model prediction accuracy is typically performed on simulated images and often focuses on detecting vacancies of relatively heavier atoms like Se (Z = 34). In Figure~\ref{fig:4}e, our data is derived from detecting lighter atom vacancies, sulfur (Z = 16), in real experimental images. This further demonstrates the robust adaptability and high precision of methods involving fine-tuning models using a single image. To broaden the application scope of our method, we have deployed the optimized models on a public website for testing (URL provided in the SI). Three main functions are presented in Figure~\ref{fig:4}f-h.

\begin{figure}[htb]
		\centering
            \includegraphics[width=0.98\textwidth]{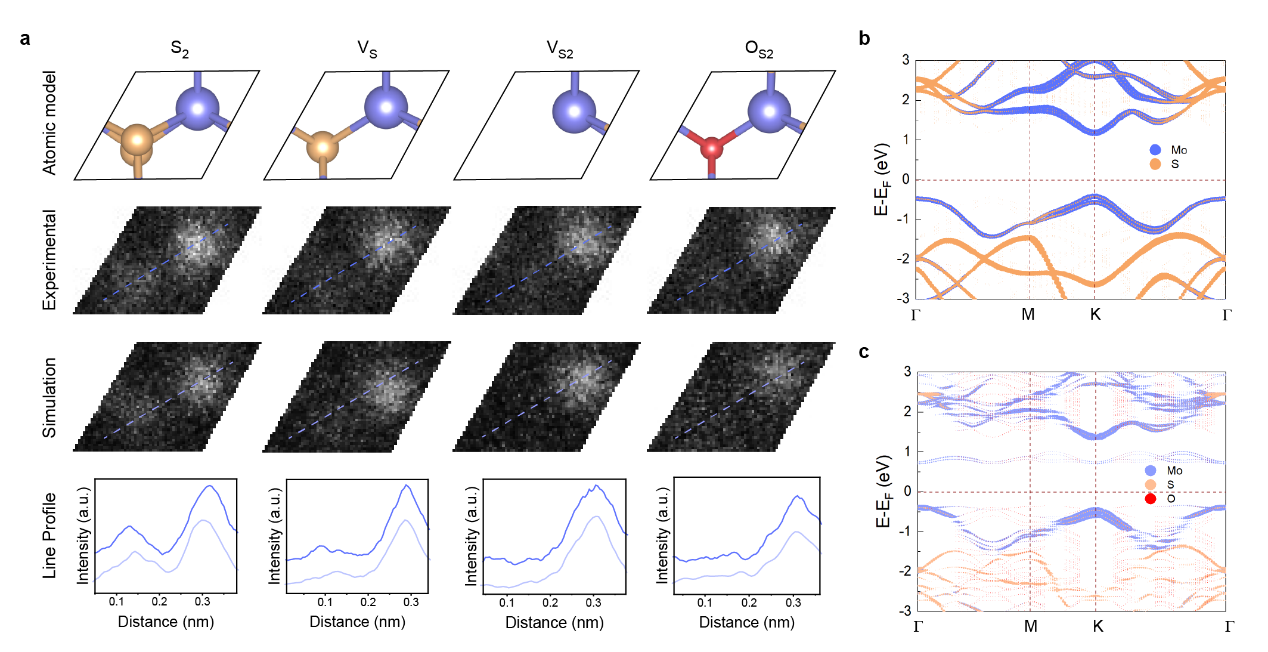}
		\caption{\textbf{HAADF-STEM images and band structures of defective monolayer $MoS_2$.} (a) Atomic model, experimental and simulated STEM images, and corresponding line profiles (from top to bottom) of pristine $MoS_2$, $V_S$, $V_{S2}$, and $O_{S2}$ (from left to right). Blue and light blue represent line profiles of experimental and simulated images, respectively. b, c) The calculated band structure of pristine (b) and O doping $MoS_2$ (c).}
		\label{fig:5}
\end{figure}

Reliable and statistical identification of oxygen atoms in transition metal dichalcogenide (TMD) lattice with atom precision is rather challenging, as the weak signals from the light O would be easily overwhelmed by the scanning noise and strong background signals from the heavy metal atoms. The use of fine-tune trained FCN enabled this hard appreciation because of its precise confidence selection. The similarity of experimental and noised-simulated STEM images of $V_S$, $V_{S2}$, $O_{S2}$ (O substituting $S_2$ site), and pristine structure ($S_2$) is displayed in Figure~\ref{fig:5}a. As suggested by intensity line profile analysis, the dim contrast at the S vacancy site in the STEM-HAADF image is due to either $V_S$ or $O_{S2}$, since the contrast difference between them is almost indiscernible. As point defects in TMDs significantly alter their band structures, we calculated the band structure of $MoS_2$ with an O substitution defect, where a pair of S atoms were replaced by one oxygen atom in a 3 × 3 × 1 supercell. As depicted in Figure~\ref{fig:5}c, our findings reveal that the introduction of O induces notable in-gap states. Upon O doping, the energy of the top valence band at the K point is marginally lower than that at the $\Gamma$ point, thereby transforming the material from a direct bandgap to an indirect bandgap semiconductor. Furthermore, O doping leads to an increase in the bandgap at the K point, elevating it from the original 1.60 (Figure~\ref{fig:5}b) to 1.82 eV (Figure~\ref{fig:5}c). 

\begin{figure}[htb]
		\centering
            \includegraphics[width=0.98\textwidth]{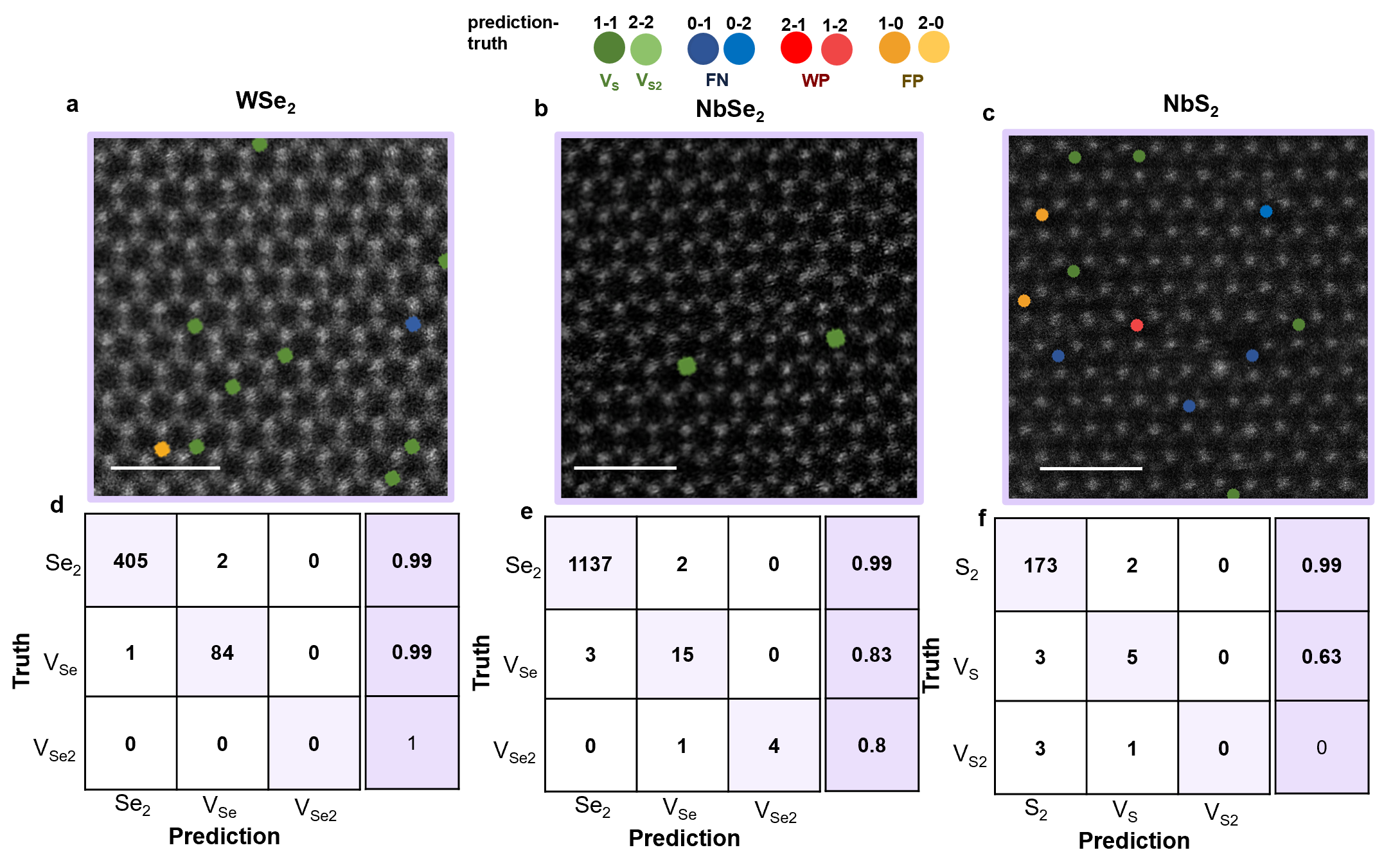}
		\caption{\textbf{Defects prediction on the experimental image of $WSe_2$, $NbS_2$, and $TaSe_2$.} (a-c) Experimental STEM images of $WSe_2$, $NbS_2$, and $TaSe_2$ inserted with labeled and predicted defects. (d-f) Confusion matrix was obtained by matching the ground truth and the predictions identified by U-Net trained on fine-tuned U-GAT-IT processed simulated images. Scale bars: 1 nm in (a-c).}
		\label{fig:6}
\end{figure}

The generality of the single-image trained defect identification model is verified by investigating three more TMD monolayer samples, $WSe_2$, $TaSe_2$, and $NbS_2$, which possess the same minimal unit structure but differ in atomic column image contrast. Figure~\ref{fig:6} summarizes the results obtained from the U-Net model trained on realistic datasets generated by a dynamically evolved model from corresponding experimental STEM images. We found that the single-image trained U-Net model is remarkably robust across various TMD materials, accurately identifying specific defects, as demonstrated in Figure~\ref{fig:6}. Notably, the accuracy of model prediction for transition metal sulfides ( $MoS_2$ and $NbS_2$) almost matches transition metal selenides ( $WSe_2$ and $TaSe_2$), implying segmenting primitive cells can be harnessed to improve the veracity of the defect identification model, and demonstrating the optimistic generalizability and extensive applicability of our method.

\section{Conclusion}
In summary, by segmenting images designated for defect detection and their corresponding simulated images into unit cell images, we have innovated in constructing a dynamically evolved U-GAT-IT model using only one single experimental image. By training the U-Net models utilizing the simulated images generated by U-GAT-IT model extracting noise precisely from that specific STEM image, we achieved a big leap in defect identification accuracy using $MoS_2$ as an example. Due to the training set providing the most accurate description of the test image, it is unsurprising that U-Net achieved remarkable high precision in defect identification. Based on the model's precise defect identification confidence, we accurately identified chalcogen vacancies, especially sulfur vacancies and even discovered oxygen doping during the training process and, after validation, expanded the recognition capabilities to identify such defects. Experiments on other types of samples have also been conducted, maintaining a high level of accuracy, and demonstrating the potential of this method for defect identification in 2D samples with various structure features. We believe that following this approach, it is possible to integrate the identification of multiple configurations of defects, such as point,line defects, etc, across different samples, presenting great opportunities to massively and intelligently learn atomic-scale images.

\section{Methods}

\subsection{DFT calculations}

The first-principles calculations were performed using the density functional theory as indicated in the Vienna ab initio simulation package. The projector augmented-wave method~\cite{2} with a 500 eV plane-wave kinetic cut-off energy was used. The generalized gradient approximation of the Perdew–Burke–Ernzerhof ~\cite{3} form was used for exchange-correlation functions. A 7×7×1 k-points mesh with a 3×3×1 supercell was used for the optimization and self-consistent calculations. The atomic positions and lattice parameters were fully optimized until forces on each atom were less than 0.001 eV/Å, and the energy convergence criterion was set at $10^{-6}$ eV for the electronic self-consistent loop. A 15 Å vacuum layer was set in the z-direction considering the two-dimensional periodicity in the plane.

\subsection{Implement Details of U-GAT-IT}

Our method is implemented using Pytorch (https://https://pytorch.org/) Python deep-learning library. Inspired by UGATIT, our generator network adopts an encoder-decoder architecture, which consists of residual blocks with instance normalization and adaptive instance layer normalization. We add a Fourier-space image discriminator to supervise the generated high-frequency noise in Fourier-space~\cite{24}. For the real-space image discriminator, we employ PatchGAN with spectral normalization, while for the Fourier-image discriminator, we use vanilla PatchGAN. To achieve the generative task, we utilize adversarial loss. Additionally, we incorporate cycle loss to preserve structural information, identity loss to match the input-output gray distribution, and CAM loss to exploit the difference between simulated and experimental data. The loss weights for these components are set to 1, 5, 10 and 100 respectively.

\subsection{U-GAT-IT training}

All training procedures utilize the Adam optimizer with a batch size of 2. The learning rate remains constant until half of the training iterations and then linearly decays to zero. In the pre-training stage, we train our network using cells cropped from simulated and experimental stem images. We set the initial learning rate to 0.0001 and the maximum iteration to 10,000. In the fine-tuning stage, we further refine the pre-trained model using the cells cropped from simulated images and the experimental image to be analyzed. We use a smaller initial learning rate of 0.00001 and train for 3,000 iterations to prevent overfitting.

\subsection{Dataset}

Our pre-trained dataset comprises 60 simulated images with varying pixel sizes and 30 experimental images under diverse conditions, including different pixel sizes, carbon contamination, scanning distortion, and low signal-to-noise ratios. In total, we extracted approximately 20,000 segmented unit-cell-level patches (four cells per patch) from both the simulated and experimental datasets, respectively. Each patch was resized to 256x256 and individually normalized to a range of [-1, 1] for training purposes. During the fine-tune stage, the number of extractable unit-cell level patches depends on the experimental data to be analyzed, roughly ranging from several hundred to just over a thousand. We use a comparable amount of simulated data to perform fine-tuning, hoping that the simulated data can learn the noise distribution of the experimental data. 

\subsection{U-Net training}

We employed a U-Net architecture featuring a sequence of convolutional blocks, each constituted of two convolutional layers with batch normalization and ReLU activation. The network begins with an input channel, progressively increasing the feature depth through the encoder stages (64, 128, 256, 512). Following the encoder, a middle block further amplifies the feature channels to 1024. The decoder mirrors this setup but in reverse, gradually reducing the feature channels (512, 256, 128, 64) and incorporating up-convolutions to restore spatial dimensions. Each decoder stage receives additional context through skip connections from the encoder. The final output is generated through a 1x1 convolution that maps the features to the desired number of classes.
We employed the Adam optimizer with a hybrid loss function combining cross-entropy loss, dice loss, and focal loss. The weights for no defect, single sulfur defect, and double sulfur defect were set at 0.2, 0.4, and 0.4 respectively. We randomly select 1000 images outputted by the U-GAT-IT model in size 256 x 256 along with corresponding labels, which are used as inputs for fine-tuned training. These are divided into a training set and a validation set with a proportion of 9:1. We train for 10 epochs based on our pretrained model.

\section*{Acknowledgments}

X.Z. thanks the Fundamental Research Funds for the Central Universities, the National Natural Science Foundation of China (grant no.52273279), and the Beijing Natural Science Foundation (No. Z220020).

\section*{Author Contributions}

\# These authors contributed equally to this work. X.Z. and Y.H. conceived and designed the experiments. Under supervision by X.Z. and J.Z., K.L., and X.H. performed sample preparation, STEM imaging, and images analysis. Under supervision by X.Z. and Y.H., Y.M., X.C., J.L., and W.H. generated simulated images and cropped unit images for U-GAT-IT and U-Net training. Under supervision by Y.H., G.K., L.Y., and L.Z., Y.M., X.C., J.L., and Z.X. constructed the U-Net models. Y.H., Y.M., X.C., J.L. contributed to the U-GAT-IT and U-Net performance evaluation. J.-Y.Y. performed DFT calculations. All the authors discussed the results and contributed to preparing the manuscript.

\section*{Competing interests}

The authors declare no competing interests.

%%%%%%%%% REFERENCES
{\small
\bibliographystyle{unsrt}
\bibliography{refs}
}

\newpage
\newpage
\appendix
\onecolumn

\end{document}